\documentclass[
prl,showpacs,superscriptaddress,numerical,amsmath,amssymb,floatfix,
reprint,
]{revtex4-2}
\usepackage{xcolor}
\usepackage[
	pdffitwindow=true,
	colorlinks=true,
	frenchlinks=false,
        linkcolor=blue,
	anchorcolor=blue,
        citecolor=blue,
        filecolor=blue,
        urlcolor=blue,
        bookmarks=true,
        bookmarksopen=true,
	bookmarksnumbered=true,
        bookmarksopenlevel=1,
        plainpages=false,
	pdfpagelayout=TwoPageLeft,
        pdfpagelabels=true,
	breaklinks
]{hyperref}
\usepackage[per-mode=symbol,separate-uncertainty]{siunitx}
\usepackage{graphicx}
\usepackage{dcolumn}
\usepackage{bm}
\usepackage[english]{babel}
\usepackage{color}
\usepackage{todonotes}
\usepackage{hyperref}
\usepackage{units}

\newcommand{\Imod}{I_\mathrm{dc}^\mathrm{mod}}
\newcommand{\Icrit}{I_\mathrm{crit}^\mathrm{mod}}

\begin{document}
\preprint{AIP/123-QED}
\title[]{Magnon transport in $\mathrm{\mathbf{Y_3Fe_5O_{12}}}$/Pt nanostructures with reduced effective magnetization}

\author{J.~G{\"u}ckelhorn}
\email[]{janine.gueckelhorn@wmi.badw.de}
\affiliation{Walther-Mei{\ss}ner-Institut, Bayerische Akademie der Wissenschaften, 85748 Garching, Germany}
\affiliation{Physik-Department, Technische Universit\"{a}t M\"{u}nchen, 85748 Garching, Germany}
\author{T.~Wimmer}
\affiliation{Walther-Mei{\ss}ner-Institut, Bayerische Akademie der Wissenschaften, 85748 Garching, Germany}
\affiliation{Physik-Department, Technische Universit\"{a}t M\"{u}nchen, 85748 Garching, Germany}
\author{M.~M{\"u}ller}
\affiliation{Walther-Mei{\ss}ner-Institut, Bayerische Akademie der Wissenschaften, 85748 Garching, Germany}
\affiliation{Physik-Department, Technische Universit\"{a}t M\"{u}nchen, 85748 Garching, Germany}
\author{S.~Gepr{\"a}gs}
\affiliation{Walther-Mei{\ss}ner-Institut, Bayerische Akademie der Wissenschaften, 85748 Garching, Germany}
\author{H.~Huebl}
\affiliation{Walther-Mei{\ss}ner-Institut, Bayerische Akademie der Wissenschaften, 85748 Garching, Germany}
\affiliation{Physik-Department, Technische Universit\"{a}t M\"{u}nchen, 85748 Garching, Germany}
\affiliation{Munich Center for Quantum Science and Technology (MCQST), D-80799 M\"{u}nchen, Germany}
\author{R.~Gross}
\affiliation{Walther-Mei{\ss}ner-Institut, Bayerische Akademie der Wissenschaften, 85748 Garching, Germany}
\affiliation{Physik-Department, Technische Universit\"{a}t M\"{u}nchen, 85748 Garching, Germany}
\affiliation{Munich Center for Quantum Science and Technology (MCQST), D-80799 M\"{u}nchen, Germany}
\author{M.~Althammer}
\email[]{matthias.althammer@wmi.badw.de}
\affiliation{Walther-Mei{\ss}ner-Institut, Bayerische Akademie der Wissenschaften, 85748 Garching, Germany}
\affiliation{Physik-Department, Technische Universit\"{a}t M\"{u}nchen, 85748 Garching, Germany}

\date{\today}

\pacs{}
\keywords{}

\begin{abstract}
For applications making use of magnonic spin currents damping effects, which decrease the spin conductivity, have to be minimized. We here investigate the magnon transport in an yttrium iron garnet thin film with strongly reduced effective magnetization. We show that in a three-terminal device the effective magnon conductivity can be increased by a factor of up to six by a current applied to a modulator electrode, which generates damping compensation above a threshold current. Moreover, we find a linear dependence of this threshold current on the applied magnetic field. We can explain this behavior by the reduced effective magnetization and the associated nearly circular magnetization precession.

\end{abstract}
\maketitle
Pure spin currents carried by magnons, the elementary excitations of the spin system in magnetically ordered insulators (MOIs), have drawn much attention due to their potential applications in information processing at a low dissipation level\,\cite{Chumak2015magnon,Chumak_2017,Nakata_2017,LudoTransistor}. 
The MOI yttrium iron garnet ($\mathrm{Y_3Fe_5O_{12}}$, YIG) is a promising candidate for hosting efficient magnon based spin transport due to its low Gilbert damping parameter even in nanometer-thin films\,\cite{CHEREPANOV199381} and its correspondingly large magnon propagation length\,\cite{Saitoh2010,CornelissenMMR,Yu2016PropagationLength,Liu2018}.
Amongst the device concepts enabling logic operation with magnonic spin currents, transistor-inspired devices and even logic gates have been demonstrated\,\cite{LudoTransistor,Wimmer2019spin,JanineACvsDC,BartvanWees_Transistor_Py,KathrinLogik}.
Such transistor-like device concepts generally rely on spin-transfer torque for spin current generation.
The latter
can be realized in bilayers consisting of MOIs and heavy metals with strong spin-orbit coupling via the spin Hall effect (SHE)\,\cite{HirschSHE,Dyakonov,CornelissenMMR,SchlitzMMR,Casanova2016,GilesSSE,VanWeesThickness,Althammer2018,althammer2021allelectrical}. 
The magnon transport in the MOI can be controlled via an electrical charge current,
and the resulting effect is typically represented by 
a change in the effective magnon conductivity\,\cite{LudoTransistor,Wimmer2019spin}.
At a certain threshold current the injected magnons can even counteract the magnetization damping, which results in an abrupt increase of the effective magnon conductivity. The present understanding is that this threshold effect scales with the saturation magnetization $M_\mathrm{s}$ and the magnetic anisotropy of the materials\,\cite{MagnonBEC,Wimmer2019spin}. 
This warrants to explore the impact of these parameters on controlling the magnon conductivity in MOIs, which has not been pursued so far to the best of our knowledge.

In this Letter, 
we investigate the diffusive magnon transport in MOIs with significant perpendicular magnetic anisotropy fields $H_\mathrm{k}$ and reduced $M_\mathrm{s}$. To this end, we biaxially strain the YIG thin film hosting the magnons by
growing YIG on yttrium scandium gallium garnet ($\mathrm{Y_3Sc_2Ga_3O_{12}}$, YSGG).
Our films exhibit low Gilbert damping comparable to YIG thin films grown on lattice-matched substrates. By investigating the magnon transport
in three-terminal devices, we find that the threshold current, which defines the onset of the regime with compensated damping, depends linearly on the applied magnetic field. Moreover, we can corroborate the expected scaling with the effective magnetization of the MOI.  

\begin{figure}
	\includegraphics[]{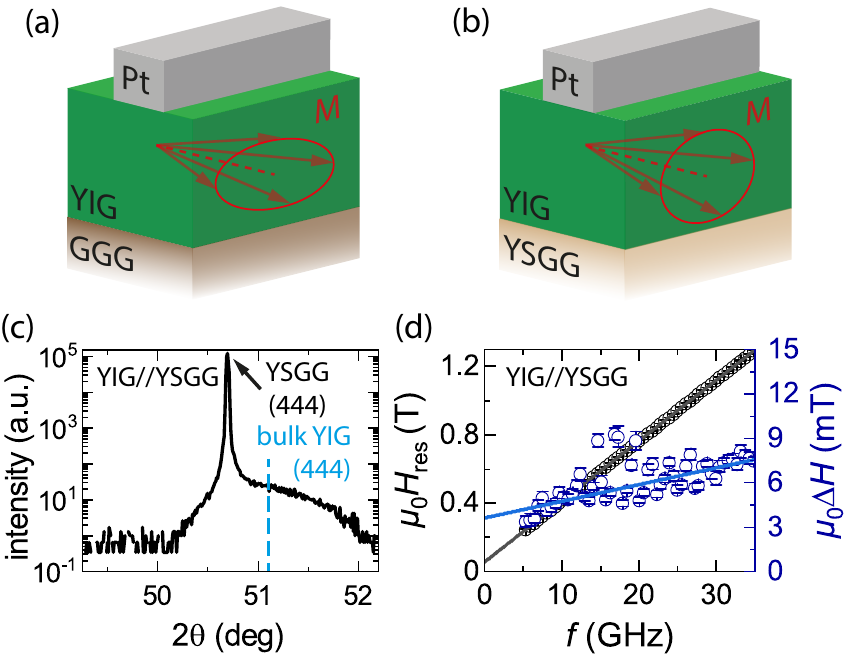}
	\caption[]{(a) Sketch of the ellipticity of the magnetization precession in YIG thin films grown on lattice-matched GGG. (b) In biaxially strained YIG thin films grown on YSGG, the ellipticity is minimized due to the reduced effective magnetization. (c) X-ray diffraction of a $\SI{12.3}{\nano\meter}$ thick YIG film grown on a (111)-oriented YSGG substrate. The blue vertical line marks the calculated $2\theta$-position of the (444) reflection of bulk YIG. (d) Resonance field $H_\mathrm{res}$ and linewidth $\Delta H$ extracted from FMR measurements of the YIG film on YSGG as a function of frequency. Via a Kittel fit (gray line) we extract $\mu_0M_\mathrm{eff}=\SI{56\pm0.2}{\milli\tesla}$ and from a linear fit to the linewidth (blue line) we obtain $\mu_0\delta H = \SI{3.6\pm0.4}{\milli\tesla}$ and $\alpha_\mathrm{G} = \num{1.5\pm0.2 e-3}$.}
	\label{fig:fig1Charactization}
\end{figure}

Our experimental approach utilized to enhance the magnon-based spin conductivity is based on the minimization of the ellipticity of the magnetization precession.
As sketched in Fig.\,\ref{fig:fig1Charactization}(a), YIG thin films grown on the lattice-matched substrate gadolinium gallium garnet ($\mathrm{Gd_3Ga_5O_{12}}$, GGG) exhibit a finite in-plane effective magnetization ${M_\mathrm{eff}=M_\mathrm{s}-H_\mathrm{k}>0}$, and thus an elliptical magnetization precession trajectory with the long axis aligned in the film plane, giving rise to nonlinear damping effects via parametric pumping of higher frequency magnon modes\,\cite{SUHLparametricpump}. Recent experiments reported the minimization of the ellipticity of the magnetization precession (approaching $M_\mathrm{eff}=0$) and thereby achieved spin-orbit torque induced coherent magnetization auto-oscillations even in extended magnetic films\,\cite{EveltBiYIG,Divinsky2019}. For our experiments, we also reduce the ellipticity of the magnetization precession by reducing the effective magnetization of YIG. Approaching $M_\mathrm{eff}=0$, a circular magnetization precession is expected and, hence, nonlinear damping effects should be suppressed (cf. Fig.\,\ref{fig:fig1Charactization}(b)).
To be able to control $M_\mathrm{eff}$, we biaxially strain the $t_\mathrm{YIG}=\SI{12.3}{\nano\meter}$ thick YIG film by growing it pseudomorphically onto an YSGG substrate by pulsed laser deposition (see the Supplemental Material (SM)\,\footnote{\label{fn:SI} See Supplemental Material at [url], which contains information on the sample fabrication, measurement procedure and a derivation of the used fitting functions for the critical modulator current. It contains Ref.\,\cite{ZhangInterfaceTransparency}.} for growth details). The lattice mismatch of $\SI{0.4}{\percent}$ between YIG and YSGG induces a biaxial in-plane tensile strain in the YIG thin film. This strain can result in a strong $H_\mathrm{k}$\,\cite{Guo2019}, 
originating from the strain-induced magnetoelastic coupling\,\cite{Popova2001}.
Fig.\,\ref{fig:fig1Charactization}(c) shows the $2\theta-\omega$ x-ray diffraction scan of the thin film confirming the 
in-plane lattice strain. The substrate (444) diffraction peak is clearly visible at $2\theta=\SI{50.7}{\degree}$, while the corresponding broad film peak is shifted to larger $2\theta$ values due to the tensile strain and appears as a shoulder in the diffration pattern. Note, that the large width and low intensity of the film peak 
is due to the small film thickness. 
We magnetically characterize the strained YIG film using broadband ferromagnetic resonance (FMR) as shown in Fig.\,\ref{fig:fig1Charactization}(d). We determine the effective magnetization  $\mu_0M_\mathrm{eff} = \SI{56\pm0.2}{\milli\tesla}$ of the thin film by extracting the resonance field $\mu_0H_\mathrm{res}$ applied in the out-of-plane direction as a function of the stimulus frequency of the microwave radiation and linear fitting with the Kittel equation.
This value is about three times smaller compared to unstrained YIG films of similar thickness\,\cite{Wimmer2019spin}.
Moreover, FMR enables us to determine the Gilbert damping parameter $\alpha_\mathrm{G}$ (see Fig.\,\ref{fig:fig1Charactization}(d))\,\cite{DerivativeDivide2018}. By fitting the FMR linewidth $\Delta H$ to $\mu_0\Delta H = \mu_0\delta H + 4\pi f\alpha_\mathrm{G}/\gamma$ (blue line) with $\gamma=\frac{g\mu_\mathrm{B}}{\hbar}$ the gyromagnetic ratio with the Landé factor $g$ and Bohr's magneton $\mu_\mathrm{B}$, we obtain the inhomogenous FMR linewidth $\mu_0\delta H = \SI{3.6\pm0.4}{\milli\tesla}$ and $\alpha_\mathrm{G} = \num{1.5\pm0.2 e-3}$. Similar values for $\alpha_\mathrm{G}$ were obtained for an epitaxial high-quality YIG film grown on lattice-matched GGG under the same conditions\,\cite{Wimmer2019spin}.

\begin{figure}
	\includegraphics[]{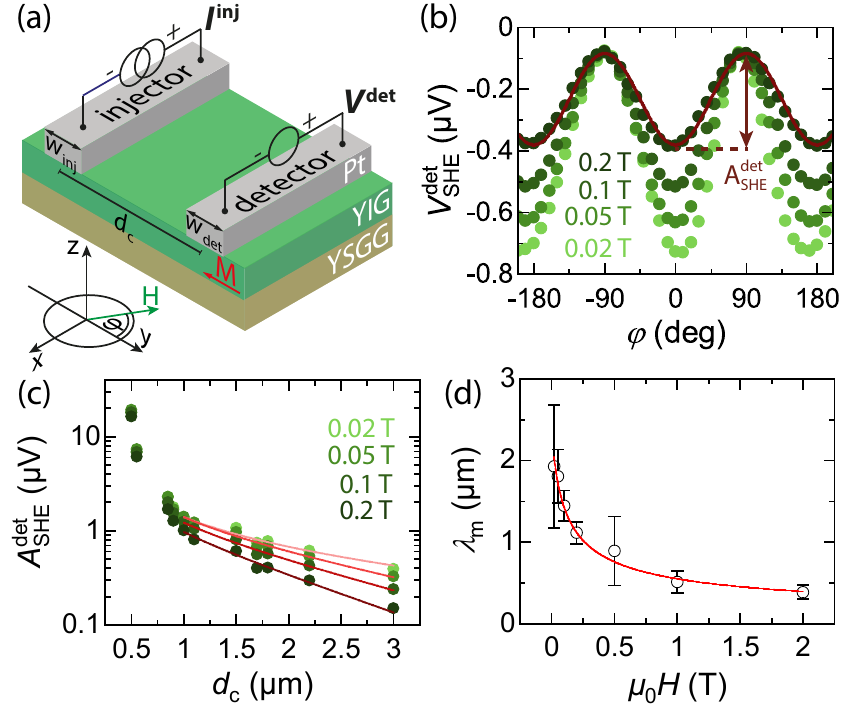}
	\caption[]{(a) Sketch of the sample configuration with the electrical connection scheme, and the coordinate system with the in-plane rotation angle $\varphi$ of the applied magnetic field $\mu_0\mathbf{H}$. (b) Detector signal  $V^\mathrm{det}_\mathrm{SHE}$ plotted versus the magnetic field orientation $\varphi$ for different magnetic field magnitudes $\mu_0H$. The red line is a fit to $A^\mathrm{det}_\mathrm{SHE}\cos^2(\varphi)$. (c) The voltage amplitudes $A^\mathrm{det}_\mathrm{SHE}$, as indicated in (b),
	plotted versus the distance $d_\mathrm{c}$ for different magnetic fields on a logarithmic scale. The red lines correspond to exponential fits to the data points for $d_\mathrm{c}\geq\SI{1}{\micro\meter}$. (d) Extracted magnon spin diffusion lengths $\lambda_\mathrm{m}$ from the exponential fits in (c) for different magnetic field magnitudes $\mu_0H$. The red line is a fit to Eq.\,(\ref{eq:DiffusionConstant}), resulting in a magnon diffusion constant $D=\SI{1.75\pm0.05 e-4}{\square\meter\per\second}$.}
	\label{fig:fig2admr}
\end{figure}

As a next step, we deposit ex-situ $\SI{5}{\nano\meter}$ thick Pt strips on top of the strained YIG film using electron beam lithography and magnetron sputtering, allowing for an all-electrical generation and detection of pure spin currents\,\cite{LudoTransistor}. With this sample, we investigate diffusive magnon transport using twin-strip structures as depicted in Fig.\,\ref{fig:fig2admr}(a). In our experiments a DC charge current $I^\mathrm{inj}=\SI{100}{\micro\ampere}$ is fed through one Pt strip (injector) to inject magnons into the YIG via the SHE. The magnons diffuse away from the injector and can then be electrically detected via the inverse SHE at the second Pt strip (detector) as a voltage signal $V^\mathrm{det}$. In our sample a constant injector width of $w_\mathrm{inj}=\SI{500}{\nano\meter}$ is used, while the detector width $w_\mathrm{det}$ and the center-to-center distance $d_\mathrm{c}$ between injector and detector is varied. Using a current reversal technique we unambiguously can assign the measured detector voltage $V^\mathrm{det}_\mathrm{SHE}$ to the magnons generated at the injector via the SHE\,\cite{SchlitzMMR,KathrinLogik}.
To characterize the magnon transport, we measure the voltage signal $V^\mathrm{det}_\mathrm{SHE}$ as a function of the magnetic field orientation $\varphi$ (cf. Fig.\,\ref{fig:fig2admr}(a)) with fixed magnitude $\mu_0H$ at a temperature of $\SI{280}{\kelvin}$. The data is shown in Fig.\,\ref{fig:fig2admr}(b) for $d_\mathrm{c}=\SI{2.2}{\micro\meter}$ and $w_\mathrm{det}=\SI{500}{\nano\meter}$. The results show the distinctive $\cos^2(\varphi)$ angular variation expected for diffusive transport of SHE-generated magnons from injector to detector\,\cite{LudoTransistor,SchlitzMMR}.
The angle dependence can be fitted with a simple $A^\mathrm{det}_\mathrm{SHE}\cos^2(\varphi)$ function, as exemplary shown for $\mu_0H=\SI{200}{\milli\tesla}$ , where $A^\mathrm{det}_\mathrm{SHE}$ corresponds to the amplitude of the SHE-induced magnon transport signal. The quantity $A^\mathrm{det}_\mathrm{SHE}$ is plotted in Fig.\,\ref{fig:fig2admr}(c) as a function of $d_\mathrm{c}$ for different $\mu_0 H$. We observe a decrease of $A_\mathrm{SHE}^\mathrm{det}$ with increasing $d_\mathrm{c}$ as expected for diffusive magnon transport: at distances shorter than the magnon diffusion length $\lambda_\mathrm{m}$, $A^\mathrm{det}_\mathrm{SHE}$ follows a $1/d_\mathrm{c}$ dependence, while for larger distances the magnon relaxation dominates and an exponential decay is observed\,\cite{CornelissenMMR,VanWeesThickness}. An exponential fit to the data 
measured 
for $d_\mathrm{c}>\SI{1}{\micro\meter}$ (red lines), allows us to extract $\lambda_\mathrm{m}$. The extracted values are shown in Fig.\,\ref{fig:fig2admr}(d) as a function of the magnetic field magnitude $\mu_0H$. The 
$\lambda_\mathrm{m}$ values are of the order of $\SI{1}{\mu\meter}$ and thus in good agreement with the values found for 
YIG films grown on lattice-matched GGG\,\cite{Wimmer2019spin}. 
To discuss the physics leading to the magnetic field dependence of $\lambda_\mathrm{m}$ in more detail, we consider the magnon relaxation rate $\Gamma_\mathrm{mr}^\mathrm{ip}$, which is given by
\begin{equation}
\Gamma_\mathrm{mr}^\mathrm{ip}=\left( \alpha_\mathrm{G}+\frac{\delta H}{2\sqrt{H(H+M_\mathrm{eff})}}\right) \gamma\mu_0\left( H+\frac{M_\mathrm{eff}}{2} \right)
\end{equation}
for an in-plane magnetized film\,\cite{HillbrandsBook}.
Taking damping contributions from inhomogenous broadening $\delta H$ into account\,\cite{collet2016generation}, the damping rate $\Gamma_\mathrm{mr}^\mathrm{ip}$ diverges for a finite positive $M_\mathrm{eff}$ at $\mu_0H = 0$.
However, in the limit of $M_\mathrm{eff}=0$, the relaxation rate is constant for $\mu_0H = 0$ and we expect a strictly linear dependence on the magnetic field. Together with $\lambda_\mathrm{m}=\sqrt{D\tau_\mathrm{m}}$ and $\tau_\mathrm{m}=1/\Gamma_\mathrm{mr}^\mathrm{ip}$ with $D$ the magnon diffusion constant and $\tau_\mathrm{m}$ the magnon 
lifetime, we can describe the magnetic field dependence of $\lambda_\mathrm{m}$ determined from the twin-strip transport measurements as
\begin{equation}
\lambda_\mathrm{m}=\sqrt{\frac{D}{\gamma\mu_0\left( \alpha_\mathrm{G}H+\frac{\delta H}{2} \right)}}\;.
\label{eq:DiffusionConstant}
\end{equation}
As shown in Fig.\,\ref{fig:fig2admr}(d), the experimental data can be well fitted by Eq.\,(\ref{eq:DiffusionConstant}). Utilizing the values obtained from the FMR measurements and neglecting the field dependence of $D$, we obtain a magnon diffusion constant of $D=\SI{1.75\pm0.05 e-4}{\square\meter\per\second}$. Similar values were obtained for YIG films on GGG, supporting the quantitative understanding of the phenomenon\,\cite{LudoPRB}.

\begin{figure}
	\includegraphics[]{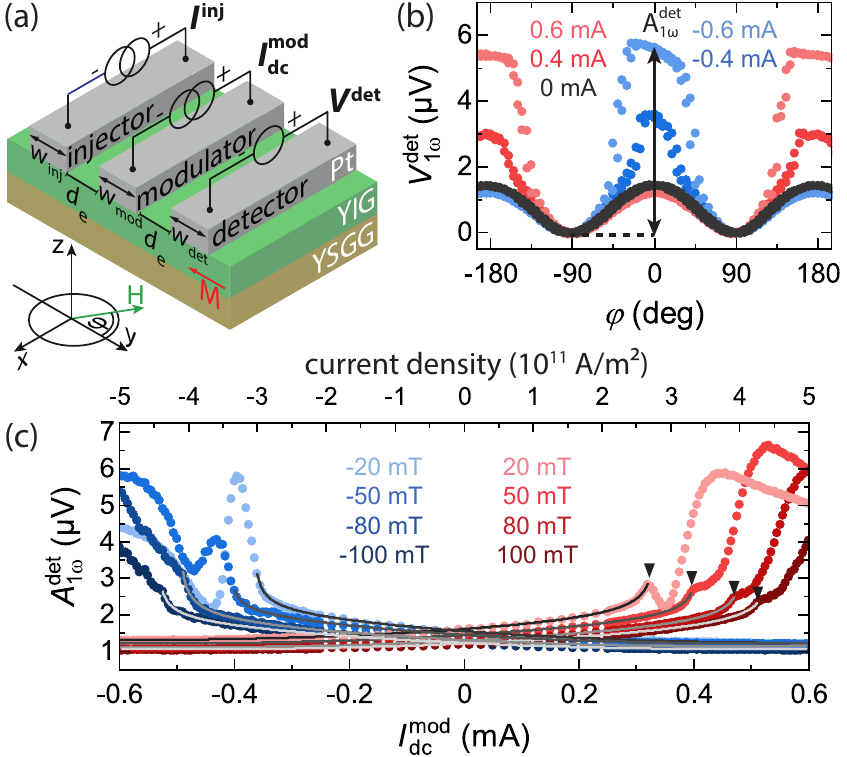}
	\caption[]{(a) Sketch of the sample configuration for a three-terminal device with the electrical connection scheme, and the coordinate system with the in-plane rotation angle $\varphi$ of the applied magnetic field $\mu_0\mathbf{H}$. (b) Detector signal  $V^\mathrm{det}_\mathrm{1\omega}$ of a structure with $d_\mathrm{e}=\SI{200}{\nano\meter}$ and $w_\mathrm{mod}=\SI{400}{\nano\meter}$ plotted versus the magnetic field orientation with constant magnitude $\mu_0H=\SI{50}{\milli\tesla}$ for various modulator currents $\Imod$. (c) The voltage amplitudes $A^\mathrm{det}_\mathrm{1\omega}$, as indicated in (b), versus the DC charge current $\Imod$. The gray lines indicate fits to Eq.\,(\ref{Eq:Amplitude_fit}) for current values below the threshold current.}
	\label{fig:fig3currsweep}
\end{figure}

Next, we turn to three-terminal devices, which allow us to manipulate the magnon transport between injector and detector via the center Pt strip acting as modulator (see Fig.\,\ref{fig:fig3currsweep}(a)). In this configuration, we apply a low-frequency ($\SI{7}{\hertz}$) charge current $I^\mathrm{inj}_\mathrm{ac}=\SI{200}{\micro\ampere}$ to the injector strip, while a constant DC charge current $I^\mathrm{mod}_\mathrm{dc}$ is applied to the modulator strip. The detector voltage $V^\mathrm{det}$ is recorded via lock-in detection, where the first harmonic voltage signal $V^\mathrm{det}_\mathrm{1\omega}$ can be assigned to the transport of magnons generated via the SHE at the injector. We measure $V^\mathrm{det}_\mathrm{1\omega}$ as a function of the magnetic field orientation $\varphi$ for
different external magnetic field magnitudes and different modulator currents $I^\mathrm{mod}_\mathrm{dc}$. Exemplary results for a structure with an edge-to-edge distance $d_\mathrm{e}=\SI{200}{\nano\meter}$ and a modulator width $w_\mathrm{mod}=\SI{400}{\nano\meter}$, while $w_\mathrm{inj}=w_\mathrm{det}=\SI{500}{\nano\meter}$ and $\mu_0H=\SI{50}{\milli\tesla}$, are plotted in Fig.\,\ref{fig:fig3currsweep}(b). For $I^\mathrm{mod}_\mathrm{dc}=0$ (black data points), $V_\mathrm{1\omega}^\mathrm{det}$ exhibits the same $\cos^2(\varphi)$ variation as in our twin-strip structures\,\cite{LudoTransistor,SchlitzMMR}. As reported previously\,\cite{Wimmer2019spin,JanineACvsDC}, we observe a significant enhancement of $V^\mathrm{det}_\mathrm{1\omega}$ at $\varphi=\pm\SI{180}{\degree}$ for $I^\mathrm{mod}_\mathrm{dc}>0$. This observation can be attributed to a magnon accumulation underneath the modulator caused by the SHE-induced magnon chemical potential and thermally generated magnons due to Joule heating. This accumulation increases the magnon conductivity, resulting in a larger voltage signal $V^\mathrm{det}_\mathrm{1\omega}$. At $\varphi = \SI{0}{\degree}$, the magnon transport signal is slightly suppressed, originating from the nearly compensation of the magnon depletion caused by the SHE by thermally generated magnons. For $I^\mathrm{mod}_\mathrm{dc}<0$, we observe a $\SI{180}{\degree}$ shifted angle dependence of the detector voltage signal, i.e. an increase at $\varphi = \SI{0}{\degree}$ and a reduction at $\varphi=\pm\SI{180}{\degree}$. This behavior is fully consistent with the assumption that there are both SHE and Joule heating contributions\,\cite{LudoTransistor,Wimmer2019spin,JanineACvsDC}.
For a more quantitative analysis, we extract the signal amplitudes $A^\mathrm{det}_\mathrm{1\omega}(+\mu_0H)$ at $\varphi = \SI{180}{\degree}$ and  $A^\mathrm{det}_\mathrm{1\omega}(-\mu_0H)$ at $\varphi = \SI{0}{\degree}$ and plot them as a function of the modulator current $I^\mathrm{mod}_\mathrm{dc}$ for different $\mu_0H$ in Fig.\,\ref{fig:fig3currsweep}(c). For $\left| \Imod\right|<\SI{0.25}{\milli\ampere}$, we observe the expected superposition of a linear and quadratic $\Imod$ dependence corresponding to SHE induced magnons and thermally generated magnons due to Joule heating, respectively \,\cite{LudoTransistor,Wimmer2019spin,JanineACvsDC}. However, for larger $\Imod$ a clear deviation from this behavior is observed. In particular, we observe a strongly increased signal amplitude $A^\mathrm{det}_\mathrm{1\omega}$. This observation can be attributed to an enhanced effective magnon conductivity underneath the modulator, which causes a strong increase of the detector signal at the same magnon injection rate at the injector. As reported previously\,\cite{Wimmer2019spin}, this enhanced magnon conductivity can be explained by the presence of a zero effective damping state generated below the modulator electrode via the SHE-mediated spin-orbit torque.
We here observe a maximum enhancement of $A^\mathrm{det}_\mathrm{1\omega}$ by a factor of 6, a twofold increase as compared to our previous experiments. This strong enhancement can be attributed to the reduction in $M_\mathrm{eff}$ and the associated circular magnetization precession. 
For the two magnetic field polarities, we observe an asymmetry for $\left| \Imod\right|>\SI{0.25}{\milli\ampere}$ in the amplitude signal $A^\mathrm{det}_\mathrm{1\omega}$. This is in stark contrast to the results obtained for YIG films on lattice-matched GGG\,\cite{Wimmer2019spin,JanineACvsDC}.
At present, we can only speculate about the detailed origin of this asymmetry. It may be related to a combination of the following aspects: (i) a misalignement of the magnetic field due to trapped flux from our 3D-vector magnet, (ii) a Joule heating induced modification of the device properties, or (iii) effects related to the crystalline-orientation of YIG.
The previously investigated YIG films were (001)-oriented\,\cite{Wimmer2019spin}, while here we use a (111)-orientation allowing us to make use of the crystalline magnetic anisotropy.

The zero effective damping state and the corresponding peak-like structure in the magnon conductivity at the threshold value $\Icrit$ was recently discussed by S. Takei\,\cite{Takei_Theory}. According to this model considerations, one can express the expected dependence of $A^\mathrm{det}_\mathrm{1\omega}$ originating from the thermal and SHE injection of magnons by
\begin{equation}
A^\mathrm{det}_\mathrm{1\omega}
\left( I^\mathrm{mod}_\mathrm{dc}\mathrm{,}\pm\mu_0H \right)
= \frac{A+B\sqrt{1\mp\nicefrac{\Imod}{\Icrit}}}{1+C\sqrt{1\mp\nicefrac{\Imod}{\Icrit}}}\;, 
\label{Eq:Amplitude_fit}
\end{equation}
where the proportionality factors account for the induced magnon conductivity and $A$, $B$, $C$ and $\Icrit$ are used as fit parameters. Note that the model is only valid up to $\Imod=\Icrit$ and we thus restrict the fit with Eq.\,(\ref{Eq:Amplitude_fit}) to this region. The fit, indicated by gray lines in Fig.\,\ref{fig:fig3currsweep}(c), reproduces well the measured data points. Although this model does not account for the amplitude asymmetry, it is well suited to extract the threshold current $\Icrit$.

\begin{figure}
	\includegraphics[]{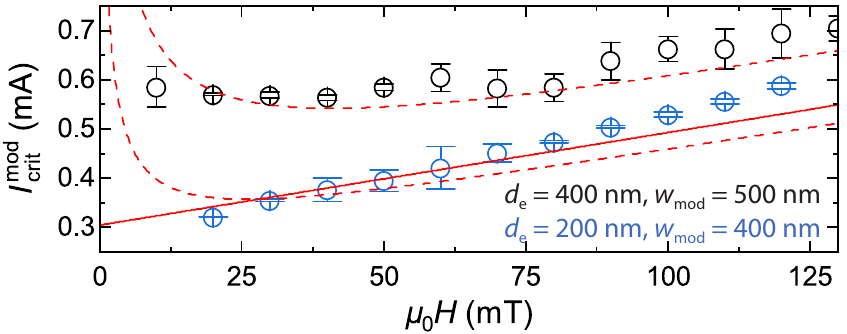}
	\caption[]{Extracted critical currents $\Icrit$, as indicated by the black triangles in Fig.\,\ref{fig:fig3currsweep}(c), as a function of the magnetic field magnitude $\mu_0 H$ (blue circles). For comparison, the black data points are taken from our previous work, where we investigate an YIG thin film grown on lattice-matched GGG substrates\,\cite{Wimmer2019spin}. The dashed lines correspond to fits to Eq.\,(\ref{eq:Icrit}) with a finite $M_\mathrm{eff}$, while the solid line is a fit to the data in the limit of $M_\mathrm{eff}=0$.}
	\label{fig:fig4Icrit}
\end{figure}

For a quantitative comparison of the strained YIG films with reduced $M_\mathrm{eff}$ and conventional YIG thin films on GGG, we rely on the dependence of $\Icrit$ with $\mu_0H$ in Fig.\,\ref{fig:fig4Icrit}. For the discussed structure (blue circles), we observe a linear increase of the critical current $\Icrit$ with applied magnetic field for $\mu_0H>\SI{20}{\milli\tesla}$. This is in contrast to the observations in Ref.\,\cite{Wimmer2019spin} (black circles), where an increase in $\Icrit$ with $\mu_0H$ is only observed for $\mu_0H>\SI{50}{\milli\tesla}$, while for $\mu_0H\leq\SI{50}{\milli\tesla}$ $\Icrit$ remains constant. 
We note that $\Icrit$ vs $\mu_0H$ was associated with damping compensation\,\cite{Wimmer2019spin}. Here, the spin injection rate due to SHE results in an interfacial spin transfer torque $\Gamma_\mathrm{ST}\propto \Imod$, which balances the intrinsic damping of the material. Hence, zero effective damping is achieved, when the condition $\Gamma_\mathrm{mr}^\mathrm{ip}=\Gamma_\mathrm{ST}$ is satisfied\,\cite{HillbrandsBook}.
In this regime, we can define the critical modulator current as\,\cite{Wimmer2019spin}
\begin{equation}
\begin{split}
\Icrit &=  \frac{\hbar}{e}\frac{\sigma_\mathrm{Pt}}{2 l_\mathrm{s}} \frac{ t_\mathrm{Pt} w_\mathrm{mod}}{\theta_\mathrm{SH} \tanh(\eta)} \left(1+4 \pi M_\mathrm{s} t_\mathrm{YIG} \frac {\alpha_\mathrm{eff}}{\hbar \gamma g_\mathrm{eff}} \right) \\  &\times \gamma\mu_0\left(H + \frac{M_\mathrm{eff}}{2}\right) \;,
\end{split}
\label{eq:Icrit}
\end{equation}
where $e$ is the elementary charge, $\theta_\mathrm{SH}$ the spin Hall angle of Pt, $\alpha_\mathrm{eff}$ the field dependent damping rate \footnote{$\alpha_\mathrm{eff}=\alpha_\mathrm{G}+\delta H\left( 2\sqrt{H(H+M_\mathrm{eff})}\right)^{-1}$} taking into account inhomogenous broadening. Furthermore, $g_\mathrm{eff}$ denotes the effective spin mixing conductance\,\footnote{
$g_\mathrm{eff}=\left[ {g^{\uparrow\downarrow} \frac{h \sigma_\mathrm{Pt}}{2e^2l_\mathrm{s}}}\right] / \left[ {g^{\uparrow\downarrow} + \frac{h\sigma_\mathrm{Pt}}{2e^2l_\mathrm{s}}} \right] $ and $\eta=t_\mathrm{Pt}/(2l_\mathrm{s})$}, which depends on the interface spin mixing conductance $g^{\uparrow\downarrow}$, the spin diffusion length $l_\mathrm{s}$, thickness $t_\mathrm{Pt}$, and electrical conductivity $\sigma_\mathrm{Pt}$ of Pt (see SM\,\ref{fn:SI}\footnotemark[1]).
The variation of $\Icrit$ with the applied magnetic field taken from Ref.\,\cite{Wimmer2019spin} is quantitatively well described by the theoretical model (dashed line). Fitting our data with Eq.\,(\ref{eq:Icrit}), we use $w_\mathrm{mod}=\SI{400}{\nano\meter}$, $M_\mathrm{s}=\SI{80}{\kilo\ampere\meter^{-1}}$ (from SQUID magnetometry measurements see SM\,\ref{fn:SI}\footnotemark[1]), $\theta_\mathrm{SH}=0.11$, $l_\mathrm{s}=\SI{1.5}{\nano\meter}$\,\cite{AlthammerSMR2013} and $\sigma_\mathrm{Pt}=\SI{2.15 e6}{({\ohm\meter})^{-1}}$. Furthermore, we use the values of $\alpha_\mathrm{G}$ and $M_\mathrm{eff}$ extracted from the FMR measurements, while $g^{\uparrow\downarrow}$ is the only free fit parameter. We observe good quantitative agreement for large magnetic field magnitudes, but find a clear deviation for $\mu_0H<\SI{40}{\milli\tesla}$.
However, if we assume $M_\mathrm{eff}\approx0$, the fit (solid line) corroborates our observed linear magnetic field dependence of $\Icrit$ in Fig.\,\ref{fig:fig4Icrit}. Moreover, the linear dependence on $\mu_0H$ is in accordance with the results by Evelt \textit{et al.}, who studied Bi:YIG thin films with PMA and nearly vanishing $M_\mathrm{eff}$\,\cite{EveltBiYIG}.
Fitting the data, we obtain  $g^{\uparrow\downarrow}=\SI{1.7\pm0.2 e19}{\meter^{-2}}$ and $g^{\uparrow\downarrow}=\SI{9.9\pm0.4 e18}{\meter^{-2}}$ in the limit of $M_\mathrm{eff}=0$, comparable to YIG/Pt structures on GGG\,\cite{Wimmer2019spin}. Deviations between fit and data are potentially caused by uncertainties in the fixed parameters, as for example $\alpha_\mathrm{G}$ and $M_\mathrm{eff}$ are determined from out-of-plane FMR.

In summary, we investigate magnon transport in YIG with strongly reduced $M_\mathrm{eff}$ induced via biaxial strain
from growth on YSGG substrates
\,\cite{Guo2019}. Performing angle-dependent measurements in twin- and three-terminal devices, we find a quantitatively similar behavior as observed for YIG films on GGG for small modulator currents $\Imod$, while differences occur above the threshold value $\Icrit$ when damping compensation is reached. Most importantly, we observe an increase of the magnon induced detector signal by a factor of about 6, which is much larger than reported previously\,\cite{Wimmer2019spin}. Another important difference is the strictly linear field dependence of $\Icrit$. This interesting observation can be attributed to the nearly vanishing $M_\mathrm{eff}$ in our film 
confirming 
the expected scaling of the threshold behavior with $M_\mathrm{s}$ and $H_\mathrm{k}$.
Our work provides an important step towards the detailed understanding of magnon transport in systems far from equilibrium and the basis for applications based on pure spin currents.

\begin{acknowledgments}
	
We gratefully acknowledge financial support from the Deutsche Forschungsgemeinschaft (DFG, German Research Foundation) under Germany's Excellence Strategy -- EXC-2111 -- 390814868 and project AL2110/2-1.
	
\end{acknowledgments}


\begin{thebibliography}{37}%
	\makeatletter
	\providecommand \@ifxundefined [1]{%
		\@ifx{#1\undefined}
	}%
	\providecommand \@ifnum [1]{%
		\ifnum #1\expandafter \@firstoftwo
		\else \expandafter \@secondoftwo
		\fi
	}%
	\providecommand \@ifx [1]{%
		\ifx #1\expandafter \@firstoftwo
		\else \expandafter \@secondoftwo
		\fi
	}%
	\providecommand \natexlab [1]{#1}%
	\providecommand \enquote  [1]{``#1''}%
	\providecommand \bibnamefont  [1]{#1}%
	\providecommand \bibfnamefont [1]{#1}%
	\providecommand \citenamefont [1]{#1}%
	\providecommand \href@noop [0]{\@secondoftwo}%
	\providecommand \href [0]{\begingroup \@sanitize@url \@href}%
	\providecommand \@href[1]{\@@startlink{#1}\@@href}%
	\providecommand \@@href[1]{\endgroup#1\@@endlink}%
	\providecommand \@sanitize@url [0]{\catcode `\\12\catcode `\$12\catcode
		`\&12\catcode `\#12\catcode `\^12\catcode `\_12\catcode `\%12\relax}%
	\providecommand \@@startlink[1]{}%
	\providecommand \@@endlink[0]{}%
	\providecommand \url  [0]{\begingroup\@sanitize@url \@url }%
	\providecommand \@url [1]{\endgroup\@href {#1}{\urlprefix }}%
	\providecommand \urlprefix  [0]{URL }%
	\providecommand \Eprint [0]{\href }%
	\providecommand \doibase [0]{https://doi.org/}%
	\providecommand \selectlanguage [0]{\@gobble}%
	\providecommand \bibinfo  [0]{\@secondoftwo}%
	\providecommand \bibfield  [0]{\@secondoftwo}%
	\providecommand \translation [1]{[#1]}%
	\providecommand \BibitemOpen [0]{}%
	\providecommand \bibitemStop [0]{}%
	\providecommand \bibitemNoStop [0]{.\EOS\space}%
	\providecommand \EOS [0]{\spacefactor3000\relax}%
	\providecommand \BibitemShut  [1]{\csname bibitem#1\endcsname}%
	\let\auto@bib@innerbib\@empty
	\bibitem [{\citenamefont {Chumak}\ \emph {et~al.}(2015)\citenamefont {Chumak},
		\citenamefont {Vasyuchka}, \citenamefont {Serga},\ and\ \citenamefont
		{Hillebrands}}]{Chumak2015magnon}%
	\BibitemOpen
	\bibfield  {author} {\bibinfo {author} {\bibfnamefont {A.~V.}\ \bibnamefont
			{Chumak}}, \bibinfo {author} {\bibfnamefont {V.~I.}\ \bibnamefont
			{Vasyuchka}}, \bibinfo {author} {\bibfnamefont {A.~A.}\ \bibnamefont
			{Serga}},\ and\ \bibinfo {author} {\bibfnamefont {B.}~\bibnamefont
			{Hillebrands}},\ }\bibfield  {title} {\bibinfo {title} {Magnon spintronics},\
	}\href {https://doi.org/10.1038/nphys3347} {\bibfield  {journal} {\bibinfo
			{journal} {Nature Physics}\ }\textbf {\bibinfo {volume} {11}},\ \bibinfo
		{pages} {453} (\bibinfo {year} {2015})}\BibitemShut {NoStop}%
	\bibitem [{\citenamefont {Chumak}\ \emph {et~al.}(2017)\citenamefont {Chumak},
		\citenamefont {Serga},\ and\ \citenamefont {Hillebrands}}]{Chumak_2017}%
	\BibitemOpen
	\bibfield  {author} {\bibinfo {author} {\bibfnamefont {A.~V.}\ \bibnamefont
			{Chumak}}, \bibinfo {author} {\bibfnamefont {A.~A.}\ \bibnamefont {Serga}},\
		and\ \bibinfo {author} {\bibfnamefont {B.}~\bibnamefont {Hillebrands}},\
	}\bibfield  {title} {\bibinfo {title} {Magnonic crystals for data
			processing},\ }\href {https://doi.org/10.1088/1361-6463/aa6a65} {\bibfield
		{journal} {\bibinfo  {journal} {Journal of Physics D: Applied Physics}\
		}\textbf {\bibinfo {volume} {50}},\ \bibinfo {pages} {244001} (\bibinfo
		{year} {2017})}\BibitemShut {NoStop}%
	\bibitem [{\citenamefont {Nakata}\ \emph {et~al.}(2017)\citenamefont {Nakata},
		\citenamefont {Simon},\ and\ \citenamefont {Loss}}]{Nakata_2017}%
	\BibitemOpen
	\bibfield  {author} {\bibinfo {author} {\bibfnamefont {K.}~\bibnamefont
			{Nakata}}, \bibinfo {author} {\bibfnamefont {P.}~\bibnamefont {Simon}},\ and\
		\bibinfo {author} {\bibfnamefont {D.}~\bibnamefont {Loss}},\ }\bibfield
	{title} {\bibinfo {title} {Spin currents and magnon dynamics in insulating
			magnets},\ }\href {https://doi.org/10.1088/1361-6463/aa5b09} {\bibfield
		{journal} {\bibinfo  {journal} {Journal of Physics D: Applied Physics}\
		}\textbf {\bibinfo {volume} {50}},\ \bibinfo {pages} {114004} (\bibinfo
		{year} {2017})}\BibitemShut {NoStop}%
	\bibitem [{\citenamefont {Cornelissen}\ \emph {et~al.}(2018)\citenamefont
		{Cornelissen}, \citenamefont {Liu}, \citenamefont {van Wees},\ and\
		\citenamefont {Duine}}]{LudoTransistor}%
	\BibitemOpen
	\bibfield  {author} {\bibinfo {author} {\bibfnamefont {L.~J.}\ \bibnamefont
			{Cornelissen}}, \bibinfo {author} {\bibfnamefont {J.}~\bibnamefont {Liu}},
		\bibinfo {author} {\bibfnamefont {B.~J.}\ \bibnamefont {van Wees}},\ and\
		\bibinfo {author} {\bibfnamefont {R.~A.}\ \bibnamefont {Duine}},\ }\bibfield
	{title} {\bibinfo {title} {Spin-current-controlled modulation of the magnon
			spin conductance in a three-terminal magnon transistor},\ }\href
	{https://doi.org/10.1103/PhysRevLett.120.097702} {\bibfield  {journal}
		{\bibinfo  {journal} {Physical Review Letters}\ }\textbf {\bibinfo {volume}
			{120}},\ \bibinfo {pages} {097702} (\bibinfo {year} {2018})}\BibitemShut
	{NoStop}%
	\bibitem [{\citenamefont {Cherepanov}\ \emph {et~al.}(1993)\citenamefont
		{Cherepanov}, \citenamefont {Kolokolov},\ and\ \citenamefont
		{L'vov}}]{CHEREPANOV199381}%
	\BibitemOpen
	\bibfield  {author} {\bibinfo {author} {\bibfnamefont {V.}~\bibnamefont
			{Cherepanov}}, \bibinfo {author} {\bibfnamefont {I.}~\bibnamefont
			{Kolokolov}},\ and\ \bibinfo {author} {\bibfnamefont {V.}~\bibnamefont
			{L'vov}},\ }\bibfield  {title} {\bibinfo {title} {The saga of {YIG}: Spectra,
			thermodynamics, interaction and relaxation of magnons in a complex magnet},\
	}\href {https://doi.org/https://doi.org/10.1016/0370-1573(93)90107-O}
	{\bibfield  {journal} {\bibinfo  {journal} {Physics Reports}\ }\textbf
		{\bibinfo {volume} {229}},\ \bibinfo {pages} {81} (\bibinfo {year}
		{1993})}\BibitemShut {NoStop}%
	\bibitem [{\citenamefont {Kajiwara}\ \emph {et~al.}(2010)\citenamefont
		{Kajiwara}, \citenamefont {Harii}, \citenamefont {Takahashi}, \citenamefont
		{Ohe}, \citenamefont {Uchida}, \citenamefont {Mizuguchi}, \citenamefont
		{Umezawa}, \citenamefont {Kawai}, \citenamefont {Ando}, \citenamefont
		{Takanashi}, \citenamefont {Maekawa},\ and\ \citenamefont
		{Saitoh}}]{Saitoh2010}%
	\BibitemOpen
	\bibfield  {author} {\bibinfo {author} {\bibfnamefont {Y.}~\bibnamefont
			{Kajiwara}}, \bibinfo {author} {\bibfnamefont {K.}~\bibnamefont {Harii}},
		\bibinfo {author} {\bibfnamefont {S.}~\bibnamefont {Takahashi}}, \bibinfo
		{author} {\bibfnamefont {J.}~\bibnamefont {Ohe}}, \bibinfo {author}
		{\bibfnamefont {K.}~\bibnamefont {Uchida}}, \bibinfo {author} {\bibfnamefont
			{M.}~\bibnamefont {Mizuguchi}}, \bibinfo {author} {\bibfnamefont
			{H.}~\bibnamefont {Umezawa}}, \bibinfo {author} {\bibfnamefont
			{H.}~\bibnamefont {Kawai}}, \bibinfo {author} {\bibfnamefont
			{K.}~\bibnamefont {Ando}}, \bibinfo {author} {\bibfnamefont {K.}~\bibnamefont
			{Takanashi}}, \bibinfo {author} {\bibfnamefont {S.}~\bibnamefont {Maekawa}},\
		and\ \bibinfo {author} {\bibfnamefont {E.}~\bibnamefont {Saitoh}},\
	}\bibfield  {title} {\bibinfo {title} {Transmission of electrical signals by
			spin-wave interconversion in a magnetic insulator},\ }\href
	{https://doi.org/10.1038/nature08876} {\bibfield  {journal} {\bibinfo
			{journal} {Nature}\ }\textbf {\bibinfo {volume} {464}},\ \bibinfo {pages}
		{262} (\bibinfo {year} {2010})}\BibitemShut {NoStop}%
	\bibitem [{\citenamefont {Cornelissen}\ \emph {et~al.}(2015)\citenamefont
		{Cornelissen}, \citenamefont {Liu}, \citenamefont {Duine}, \citenamefont
		{Youssef},\ and\ \citenamefont {van Wees}}]{CornelissenMMR}%
	\BibitemOpen
	\bibfield  {author} {\bibinfo {author} {\bibfnamefont {L.~J.}\ \bibnamefont
			{Cornelissen}}, \bibinfo {author} {\bibfnamefont {J.}~\bibnamefont {Liu}},
		\bibinfo {author} {\bibfnamefont {R.~A.}\ \bibnamefont {Duine}}, \bibinfo
		{author} {\bibfnamefont {J.~B.}\ \bibnamefont {Youssef}},\ and\ \bibinfo
		{author} {\bibfnamefont {B.~J.}\ \bibnamefont {van Wees}},\ }\bibfield
	{title} {\bibinfo {title} {Long-distance transport of magnon spin information
			in a magnetic insulator at room~temperature},\ }\href
	{https://doi.org/10.1038/nphys3465} {\bibfield  {journal} {\bibinfo
			{journal} {Nature Physics}\ }\textbf {\bibinfo {volume} {11}},\ \bibinfo
		{pages} {1022} (\bibinfo {year} {2015})}\BibitemShut {NoStop}%
	\bibitem [{\citenamefont {Yu}\ \emph {et~al.}(2016)\citenamefont {Yu},
		\citenamefont {d'Allivy Kelly}, \citenamefont {Cros}, \citenamefont
		{Bernard}, \citenamefont {Bortolotti}, \citenamefont {Anane}, \citenamefont
		{Brandl}, \citenamefont {Heimbach},\ and\ \citenamefont
		{Grundler}}]{Yu2016PropagationLength}%
	\BibitemOpen
	\bibfield  {author} {\bibinfo {author} {\bibfnamefont {H.}~\bibnamefont
			{Yu}}, \bibinfo {author} {\bibfnamefont {O.}~\bibnamefont {d'Allivy Kelly}},
		\bibinfo {author} {\bibfnamefont {V.}~\bibnamefont {Cros}}, \bibinfo {author}
		{\bibfnamefont {R.}~\bibnamefont {Bernard}}, \bibinfo {author} {\bibfnamefont
			{P.}~\bibnamefont {Bortolotti}}, \bibinfo {author} {\bibfnamefont
			{A.}~\bibnamefont {Anane}}, \bibinfo {author} {\bibfnamefont
			{F.}~\bibnamefont {Brandl}}, \bibinfo {author} {\bibfnamefont
			{F.}~\bibnamefont {Heimbach}},\ and\ \bibinfo {author} {\bibfnamefont
			{D.}~\bibnamefont {Grundler}},\ }\bibfield  {title} {\bibinfo {title}
		{Approaching soft x-ray wavelengths in nanomagnet-based microwave
			technology},\ }\href {https://doi.org/10.1038/ncomms11255} {\bibfield
		{journal} {\bibinfo  {journal} {Nature Communications}\ }\textbf {\bibinfo
			{volume} {7}},\ \bibinfo {pages} {11255} (\bibinfo {year}
		{2016})}\BibitemShut {NoStop}%
	\bibitem [{\citenamefont {Liu}\ \emph {et~al.}(2018)\citenamefont {Liu},
		\citenamefont {Chen}, \citenamefont {Liu}, \citenamefont {Heimbach},
		\citenamefont {Yu}, \citenamefont {Xiao}, \citenamefont {Hu}, \citenamefont
		{Liu}, \citenamefont {Chang}, \citenamefont {Stueckler}, \citenamefont {Tu},
		\citenamefont {Zhang}, \citenamefont {Zhang}, \citenamefont {Gao},
		\citenamefont {Liao}, \citenamefont {Yu}, \citenamefont {Xia}, \citenamefont
		{Lei}, \citenamefont {ZHAO},\ and\ \citenamefont {Wu}}]{Liu2018}%
	\BibitemOpen
	\bibfield  {author} {\bibinfo {author} {\bibfnamefont {C.}~\bibnamefont
			{Liu}}, \bibinfo {author} {\bibfnamefont {J.}~\bibnamefont {Chen}}, \bibinfo
		{author} {\bibfnamefont {T.}~\bibnamefont {Liu}}, \bibinfo {author}
		{\bibfnamefont {F.}~\bibnamefont {Heimbach}}, \bibinfo {author}
		{\bibfnamefont {H.}~\bibnamefont {Yu}}, \bibinfo {author} {\bibfnamefont
			{Y.}~\bibnamefont {Xiao}}, \bibinfo {author} {\bibfnamefont {J.}~\bibnamefont
			{Hu}}, \bibinfo {author} {\bibfnamefont {M.}~\bibnamefont {Liu}}, \bibinfo
		{author} {\bibfnamefont {H.}~\bibnamefont {Chang}}, \bibinfo {author}
		{\bibfnamefont {T.}~\bibnamefont {Stueckler}}, \bibinfo {author}
		{\bibfnamefont {S.}~\bibnamefont {Tu}}, \bibinfo {author} {\bibfnamefont
			{Y.}~\bibnamefont {Zhang}}, \bibinfo {author} {\bibfnamefont
			{Y.}~\bibnamefont {Zhang}}, \bibinfo {author} {\bibfnamefont
			{P.}~\bibnamefont {Gao}}, \bibinfo {author} {\bibfnamefont {Z.}~\bibnamefont
			{Liao}}, \bibinfo {author} {\bibfnamefont {D.}~\bibnamefont {Yu}}, \bibinfo
		{author} {\bibfnamefont {K.}~\bibnamefont {Xia}}, \bibinfo {author}
		{\bibfnamefont {n.}~\bibnamefont {Lei}}, \bibinfo {author} {\bibfnamefont
			{W.}~\bibnamefont {ZHAO}},\ and\ \bibinfo {author} {\bibfnamefont
			{M.}~\bibnamefont {Wu}},\ }\bibfield  {title} {\bibinfo {title}
		{Long-distance propagation of short-wavelength spin waves},\ }\href
	{https://doi.org/10.1038/s41467-018-03199-8} {\bibfield  {journal} {\bibinfo
			{journal} {Nature Communications}\ }\textbf {\bibinfo {volume} {9}},\
		\bibinfo {pages} {738} (\bibinfo {year} {2018})}\BibitemShut {NoStop}%
	\bibitem [{\citenamefont {Wimmer}\ \emph {et~al.}(2019)\citenamefont {Wimmer},
		\citenamefont {Althammer}, \citenamefont {Liensberger}, \citenamefont
		{Vlietstra}, \citenamefont {Gepr\"ags}, \citenamefont {Weiler}, \citenamefont
		{Gross},\ and\ \citenamefont {Huebl}}]{Wimmer2019spin}%
	\BibitemOpen
	\bibfield  {author} {\bibinfo {author} {\bibfnamefont {T.}~\bibnamefont
			{Wimmer}}, \bibinfo {author} {\bibfnamefont {M.}~\bibnamefont {Althammer}},
		\bibinfo {author} {\bibfnamefont {L.}~\bibnamefont {Liensberger}}, \bibinfo
		{author} {\bibfnamefont {N.}~\bibnamefont {Vlietstra}}, \bibinfo {author}
		{\bibfnamefont {S.}~\bibnamefont {Gepr\"ags}}, \bibinfo {author}
		{\bibfnamefont {M.}~\bibnamefont {Weiler}}, \bibinfo {author} {\bibfnamefont
			{R.}~\bibnamefont {Gross}},\ and\ \bibinfo {author} {\bibfnamefont
			{H.}~\bibnamefont {Huebl}},\ }\bibfield  {title} {\bibinfo {title} {Spin
			transport in a magnetic insulator with zero effective damping},\ }\href
	{https://doi.org/10.1103/PhysRevLett.123.257201} {\bibfield  {journal}
		{\bibinfo  {journal} {Physical Review Letters}\ }\textbf {\bibinfo {volume}
			{123}},\ \bibinfo {pages} {257201} (\bibinfo {year} {2019})}\BibitemShut
	{NoStop}%
	\bibitem [{\citenamefont {Gückelhorn}\ \emph {et~al.}(2020)\citenamefont
		{Gückelhorn}, \citenamefont {Wimmer}, \citenamefont {Geprägs},
		\citenamefont {Huebl}, \citenamefont {Gross},\ and\ \citenamefont
		{Althammer}}]{JanineACvsDC}%
	\BibitemOpen
	\bibfield  {author} {\bibinfo {author} {\bibfnamefont {J.}~\bibnamefont
			{Gückelhorn}}, \bibinfo {author} {\bibfnamefont {T.}~\bibnamefont {Wimmer}},
		\bibinfo {author} {\bibfnamefont {S.}~\bibnamefont {Geprägs}}, \bibinfo
		{author} {\bibfnamefont {H.}~\bibnamefont {Huebl}}, \bibinfo {author}
		{\bibfnamefont {R.}~\bibnamefont {Gross}},\ and\ \bibinfo {author}
		{\bibfnamefont {M.}~\bibnamefont {Althammer}},\ }\bibfield  {title} {\bibinfo
		{title} {Quantitative comparison of magnon transport experiments in
			three-terminal {YIG/Pt} nanostructures acquired via dc and ac detection
			techniques},\ }\href {https://doi.org/10.1063/5.0023307} {\bibfield
		{journal} {\bibinfo  {journal} {Applied Physics Letters}\ }\textbf {\bibinfo
			{volume} {117}},\ \bibinfo {pages} {182401} (\bibinfo {year}
		{2020})}\BibitemShut {NoStop}%
	\bibitem [{\citenamefont {Santos}\ \emph {et~al.}(2021)\citenamefont {Santos},
		\citenamefont {Feringa}, \citenamefont {Das}, \citenamefont {Youssef},\ and\
		\citenamefont {van Wees}}]{BartvanWees_Transistor_Py}%
	\BibitemOpen
	\bibfield  {author} {\bibinfo {author} {\bibfnamefont {O.~A.}\ \bibnamefont
			{Santos}}, \bibinfo {author} {\bibfnamefont {F.}~\bibnamefont {Feringa}},
		\bibinfo {author} {\bibfnamefont {K.~S.}~\bibnamefont {Das}}, \bibinfo {author}
		{\bibfnamefont {J.~B.}\ \bibnamefont {Youssef}},\ and\ \bibinfo {author}
		{\bibfnamefont {B.~J.}~\bibnamefont {van Wees}},\ }\bibfield  {title} {\bibinfo
		{title} {Efficient modulation of magnon conductivity in
			{${\mathrm{Y}}_{3}{\mathrm{Fe}}_{5}{\mathrm{O}}_{12}$} using anomalous spin
			hall effect of a permalloy gate electrode},\ }\href
	{https://doi.org/10.1103/PhysRevApplied.15.014038} {\bibfield  {journal}
		{\bibinfo  {journal} {Physical Review Applied}\ }\textbf {\bibinfo {volume}
			{15}},\ \bibinfo {pages} {014038} (\bibinfo {year} {2021})}\BibitemShut
	{NoStop}%
	\bibitem [{\citenamefont {Ganzhorn}\ \emph {et~al.}(2016)\citenamefont
		{Ganzhorn}, \citenamefont {Klingler}, \citenamefont {Wimmer}, \citenamefont
		{Gepr\"{a}gs}, \citenamefont {Gross}, \citenamefont {Huebl},\ and\
		\citenamefont {Goennenwein}}]{KathrinLogik}%
	\BibitemOpen
	\bibfield  {author} {\bibinfo {author} {\bibfnamefont {K.}~\bibnamefont
			{Ganzhorn}}, \bibinfo {author} {\bibfnamefont {S.}~\bibnamefont {Klingler}},
		\bibinfo {author} {\bibfnamefont {T.}~\bibnamefont {Wimmer}}, \bibinfo
		{author} {\bibfnamefont {S.}~\bibnamefont {Gepr\"{a}gs}}, \bibinfo {author}
		{\bibfnamefont {R.}~\bibnamefont {Gross}}, \bibinfo {author} {\bibfnamefont
			{H.}~\bibnamefont {Huebl}},\ and\ \bibinfo {author} {\bibfnamefont
			{S.~T.~B.}\ \bibnamefont {Goennenwein}},\ }\bibfield  {title} {\bibinfo
		{title} {Magnon-based logic in a multi-terminal {YIG}/{Pt} nanostructure},\
	}\href {https://doi.org/10.1063/1.4958893} {\bibfield  {journal} {\bibinfo
			{journal} {Applied Physics Letters}\ }\textbf {\bibinfo {volume} {109}},\
		\bibinfo {pages} {022405} (\bibinfo {year} {2016})}\BibitemShut {NoStop}%
	\bibitem [{\citenamefont {Hirsch}(1999)}]{HirschSHE}%
	\BibitemOpen
	\bibfield  {author} {\bibinfo {author} {\bibfnamefont {J.~E.}\ \bibnamefont
			{Hirsch}},\ }\bibfield  {title} {\bibinfo {title} {Spin hall effect},\ }\href
	{https://doi.org/10.1103/PhysRevLett.83.1834} {\bibfield  {journal} {\bibinfo
			{journal} {Physical Review Letters}\ }\textbf {\bibinfo {volume} {83}},\
		\bibinfo {pages} {1834} (\bibinfo {year} {1999})}\BibitemShut {NoStop}%
	\bibitem [{\citenamefont {Dyakonov}\ and\ \citenamefont
		{Perel}(1971)}]{Dyakonov}%
	\BibitemOpen
	\bibfield  {author} {\bibinfo {author} {\bibfnamefont {M.~I.}\ \bibnamefont
			{Dyakonov}}\ and\ \bibinfo {author} {\bibfnamefont {V.~I.}\ \bibnamefont
			{Perel}},\ }\bibfield  {title} {\bibinfo {title} {Possibility of orienting
			electron spins with current},\ }\href
	{http://jetpletters.ru/ps/1587/article_24366.shtml} {\bibfield  {journal}
		{\bibinfo  {journal} {Journal of Experimental and Theoretical Physics
				Letters}\ }\textbf {\bibinfo {volume} {13}},\ \bibinfo {pages} {467}
		(\bibinfo {year} {1971})}\BibitemShut {NoStop}%
	\bibitem [{\citenamefont {Goennenwein}\ \emph {et~al.}(2015)\citenamefont
		{Goennenwein}, \citenamefont {Schlitz}, \citenamefont {Pernpeintner},
		\citenamefont {Ganzhorn}, \citenamefont {Althammer}, \citenamefont {Gross},\
		and\ \citenamefont {Huebl}}]{SchlitzMMR}%
	\BibitemOpen
	\bibfield  {author} {\bibinfo {author} {\bibfnamefont {S.~T.~B.}\
			\bibnamefont {Goennenwein}}, \bibinfo {author} {\bibfnamefont
			{R.}~\bibnamefont {Schlitz}}, \bibinfo {author} {\bibfnamefont
			{M.}~\bibnamefont {Pernpeintner}}, \bibinfo {author} {\bibfnamefont
			{K.}~\bibnamefont {Ganzhorn}}, \bibinfo {author} {\bibfnamefont
			{M.}~\bibnamefont {Althammer}}, \bibinfo {author} {\bibfnamefont
			{R.}~\bibnamefont {Gross}},\ and\ \bibinfo {author} {\bibfnamefont
			{H.}~\bibnamefont {Huebl}},\ }\bibfield  {title} {\bibinfo {title} {Non-local
			magnetoresistance in {YIG/Pt} nanostructures},\ }\href
	{https://doi.org/http://dx.doi.org/10.1063/1.4935074} {\bibfield  {journal}
		{\bibinfo  {journal} {Applied Physics Letters}\ }\textbf {\bibinfo {volume}
			{107}},\ \bibinfo {eid} {172405} (\bibinfo {year} {2015})}\BibitemShut
	{NoStop}%
	\bibitem [{\citenamefont {V\'elez}\ \emph {et~al.}(2016)\citenamefont
		{V\'elez}, \citenamefont {Bedoya-Pinto}, \citenamefont {Yan}, \citenamefont
		{Hueso},\ and\ \citenamefont {Casanova}}]{Casanova2016}%
	\BibitemOpen
	\bibfield  {author} {\bibinfo {author} {\bibfnamefont {S.}~\bibnamefont
			{V\'elez}}, \bibinfo {author} {\bibfnamefont {A.}~\bibnamefont
			{Bedoya-Pinto}}, \bibinfo {author} {\bibfnamefont {W.}~\bibnamefont {Yan}},
		\bibinfo {author} {\bibfnamefont {L.~E.}\ \bibnamefont {Hueso}},\ and\
		\bibinfo {author} {\bibfnamefont {F.}~\bibnamefont {Casanova}},\ }\bibfield
	{title} {\bibinfo {title} {Competing effects at {Pt/YIG} interfaces: Spin
			hall magnetoresistance, magnon excitations, and magnetic frustration},\
	}\href {https://doi.org/10.1103/PhysRevB.94.174405} {\bibfield  {journal}
		{\bibinfo  {journal} {Physical Review B}\ }\textbf {\bibinfo {volume} {94}},\
		\bibinfo {pages} {174405} (\bibinfo {year} {2016})}\BibitemShut {NoStop}%
	\bibitem [{\citenamefont {Giles}\ \emph {et~al.}(2015)\citenamefont {Giles},
		\citenamefont {Yang}, \citenamefont {Jamison},\ and\ \citenamefont
		{Myers}}]{GilesSSE}%
	\BibitemOpen
	\bibfield  {author} {\bibinfo {author} {\bibfnamefont {B.~L.}\ \bibnamefont
			{Giles}}, \bibinfo {author} {\bibfnamefont {Z.}~\bibnamefont {Yang}},
		\bibinfo {author} {\bibfnamefont {J.~S.}\ \bibnamefont {Jamison}},\ and\
		\bibinfo {author} {\bibfnamefont {R.~C.}\ \bibnamefont {Myers}},\ }\bibfield
	{title} {\bibinfo {title} {Long-range pure magnon spin diffusion observed in
			a nonlocal spin-seebeck geometry},\ }\href
	{https://doi.org/10.1103/PhysRevB.92.224415} {\bibfield  {journal} {\bibinfo
			{journal} {Physical Review B}\ }\textbf {\bibinfo {volume} {92}},\ \bibinfo
		{pages} {224415} (\bibinfo {year} {2015})}\BibitemShut {NoStop}%
	\bibitem [{\citenamefont {Shan}\ \emph {et~al.}(2016)\citenamefont {Shan},
		\citenamefont {Cornelissen}, \citenamefont {Vlietstra}, \citenamefont
		{Ben~Youssef}, \citenamefont {Kuschel}, \citenamefont {Duine},\ and\
		\citenamefont {van Wees}}]{VanWeesThickness}%
	\BibitemOpen
	\bibfield  {author} {\bibinfo {author} {\bibfnamefont {J.}~\bibnamefont
			{Shan}}, \bibinfo {author} {\bibfnamefont {L.~J.}\ \bibnamefont
			{Cornelissen}}, \bibinfo {author} {\bibfnamefont {N.}~\bibnamefont
			{Vlietstra}}, \bibinfo {author} {\bibfnamefont {J.}~\bibnamefont
			{Ben~Youssef}}, \bibinfo {author} {\bibfnamefont {T.}~\bibnamefont
			{Kuschel}}, \bibinfo {author} {\bibfnamefont {R.~A.}\ \bibnamefont {Duine}},\
		and\ \bibinfo {author} {\bibfnamefont {B.~J.}\ \bibnamefont {van Wees}},\
	}\bibfield  {title} {\bibinfo {title} {Influence of yttrium iron garnet
			thickness and heater opacity on the nonlocal transport of electrically and
			thermally excited magnons},\ }\href
	{https://doi.org/10.1103/PhysRevB.94.174437} {\bibfield  {journal} {\bibinfo
			{journal} {Physical Review B}\ }\textbf {\bibinfo {volume} {94}},\ \bibinfo
		{pages} {174437} (\bibinfo {year} {2016})}\BibitemShut {NoStop}%
	\bibitem [{\citenamefont {Althammer}(2018)}]{Althammer2018}%
	\BibitemOpen
	\bibfield  {author} {\bibinfo {author} {\bibfnamefont {M.}~\bibnamefont
			{Althammer}},\ }\bibfield  {title} {\bibinfo {title} {Pure spin currents in
			magnetically ordered insulator/normal metal heterostructures},\ }\href
	{https://doi.org/10.1088/1361-6463/aaca89} {\bibfield  {journal} {\bibinfo
			{journal} {Journal of Physics D: Applied Physics}\ }\textbf {\bibinfo
			{volume} {51}},\ \bibinfo {pages} {313001} (\bibinfo {year}
		{2018})}\BibitemShut {NoStop}%
	\bibitem [{\citenamefont {Althammer}(2021)}]{althammer2021allelectrical}%
	\BibitemOpen
	\bibfield  {author} {\bibinfo {author} {\bibfnamefont {M.}~\bibnamefont
			{Althammer}},\ }\href@noop {} {\bibinfo {title} {All-electrical magnon
			transport experiments in magnetically ordered insulators}} (\bibinfo {year}
	{2021}),\ \Eprint {https://arxiv.org/abs/2103.08996} {arXiv:2103.08996}
	\BibitemShut {NoStop}%
	\bibitem [{\citenamefont {Bender}\ \emph {et~al.}(2014)\citenamefont {Bender},
		\citenamefont {Duine}, \citenamefont {Brataas},\ and\ \citenamefont
		{Tserkovnyak}}]{MagnonBEC}%
	\BibitemOpen
	\bibfield  {author} {\bibinfo {author} {\bibfnamefont {S.~A.}\ \bibnamefont
			{Bender}}, \bibinfo {author} {\bibfnamefont {R.~A.}\ \bibnamefont {Duine}},
		\bibinfo {author} {\bibfnamefont {A.}~\bibnamefont {Brataas}},\ and\ \bibinfo
		{author} {\bibfnamefont {Y.}~\bibnamefont {Tserkovnyak}},\ }\bibfield
	{title} {\bibinfo {title} {Dynamic phase diagram of dc-pumped magnon
			condensates},\ }\href {https://doi.org/10.1103/PhysRevB.90.094409} {\bibfield
		{journal} {\bibinfo  {journal} {Physical Review B}\ }\textbf {\bibinfo
			{volume} {90}},\ \bibinfo {pages} {094409} (\bibinfo {year}
		{2014})}\BibitemShut {NoStop}%
	\bibitem [{\citenamefont {Suhl}(1957)}]{SUHLparametricpump}%
	\BibitemOpen
	\bibfield  {author} {\bibinfo {author} {\bibfnamefont {H.}~\bibnamefont
			{Suhl}},\ }\bibfield  {title} {\bibinfo {title} {The theory of ferromagnetic
			resonance at high signal powers},\ }\href
	{https://doi.org/https://doi.org/10.1016/0022-3697(57)90010-0} {\bibfield
		{journal} {\bibinfo  {journal} {Journal of Physics and Chemistry of Solids}\
		}\textbf {\bibinfo {volume} {1}},\ \bibinfo {pages} {209} (\bibinfo {year}
		{1957})}\BibitemShut {NoStop}%
	\bibitem [{\citenamefont {Evelt}\ \emph {et~al.}(2018)\citenamefont {Evelt},
		\citenamefont {Soumah}, \citenamefont {Rinkevich}, \citenamefont
		{Demokritov}, \citenamefont {Anane}, \citenamefont {Cros}, \citenamefont
		{Ben~Youssef}, \citenamefont {de~Loubens}, \citenamefont {Klein},
		\citenamefont {Bortolotti},\ and\ \citenamefont {Demidov}}]{EveltBiYIG}%
	\BibitemOpen
	\bibfield  {author} {\bibinfo {author} {\bibfnamefont {M.}~\bibnamefont
			{Evelt}}, \bibinfo {author} {\bibfnamefont {L.}~\bibnamefont {Soumah}},
		\bibinfo {author} {\bibfnamefont {A.~B.}~\bibnamefont {Rinkevich}}, \bibinfo
		{author} {\bibfnamefont {S.~O.}~\bibnamefont {Demokritov}}, \bibinfo {author}
		{\bibfnamefont {A.}~\bibnamefont {Anane}}, \bibinfo {author} {\bibfnamefont
			{V.}~\bibnamefont {Cros}}, \bibinfo {author} {\bibfnamefont {J.}~\bibnamefont
			{Ben~Youssef}}, \bibinfo {author} {\bibfnamefont {G.}~\bibnamefont
			{de~Loubens}}, \bibinfo {author} {\bibfnamefont {O.}~\bibnamefont {Klein}},
		\bibinfo {author} {\bibfnamefont {P.}~\bibnamefont {Bortolotti}},\ and\
		\bibinfo {author} {\bibfnamefont {V.~E.}~\bibnamefont {Demidov}},\ }\bibfield
	{title} {\bibinfo {title} {Emission of coherent propagating magnons by
			insulator-based spin-orbit-torque oscillators},\ }\href
	{https://doi.org/10.1103/PhysRevApplied.10.041002} {\bibfield  {journal}
		{\bibinfo  {journal} {Physical Review Applied}\ }\textbf {\bibinfo {volume}
			{10}},\ \bibinfo {pages} {041002(R)} (\bibinfo {year} {2018})}\BibitemShut
	{NoStop}%
	\bibitem [{\citenamefont {Divinskiy}\ \emph {et~al.}(2019)\citenamefont
		{Divinskiy}, \citenamefont {Urazhdin}, \citenamefont {Demokritov},\ and\
		\citenamefont {Demidov}}]{Divinsky2019}%
	\BibitemOpen
	\bibfield  {author} {\bibinfo {author} {\bibfnamefont {B.}~\bibnamefont
			{Divinskiy}}, \bibinfo {author} {\bibfnamefont {S.}~\bibnamefont {Urazhdin}},
		\bibinfo {author} {\bibfnamefont {S.}~\bibnamefont {Demokritov}},\ and\
		\bibinfo {author} {\bibfnamefont {V.}~\bibnamefont {Demidov}},\ }\bibfield
	{title} {\bibinfo {title} {Controlled nonlinear magnetic damping in spin-hall
			nano-devices},\ }\href {https://doi.org/10.1038/s41467-019-13246-7}
	{\bibfield  {journal} {\bibinfo  {journal} {Nature Communications}\ }\textbf
		{\bibinfo {volume} {10}},\ \bibinfo {pages} {5211} (\bibinfo {year}
		{2019})}\BibitemShut {NoStop}%
	\bibitem [{Note1()}]{Note1}%
	\BibitemOpen
	\bibinfo {note} {\label {fn:SI} See Supplemental Material at [url], which
		contains information on the sample fabrication, measurement procedure and a
		derivation of the used fitting functions for the critical modulator current.
		It contains Ref.\protect \,\cite {ZhangInterfaceTransparency}.}\BibitemShut
	{Stop}%
	\bibitem [{\citenamefont {Guo}\ \emph {et~al.}(2019)\citenamefont {Guo},
		\citenamefont {Wan}, \citenamefont {Zhao}, \citenamefont {Wu}, \citenamefont
		{Fang}, \citenamefont {Yan}, \citenamefont {Feng}, \citenamefont {Liu},\ and\
		\citenamefont {Han}}]{Guo2019}%
	\BibitemOpen
	\bibfield  {author} {\bibinfo {author} {\bibfnamefont {C.~Y.}\ \bibnamefont
			{Guo}}, \bibinfo {author} {\bibfnamefont {C.~H.}\ \bibnamefont {Wan}},
		\bibinfo {author} {\bibfnamefont {M.~K.}\ \bibnamefont {Zhao}}, \bibinfo
		{author} {\bibfnamefont {H.}~\bibnamefont {Wu}}, \bibinfo {author}
		{\bibfnamefont {C.}~\bibnamefont {Fang}}, \bibinfo {author} {\bibfnamefont
			{Z.~R.}\ \bibnamefont {Yan}}, \bibinfo {author} {\bibfnamefont {J.~F.}\
			\bibnamefont {Feng}}, \bibinfo {author} {\bibfnamefont {H.~F.}\ \bibnamefont
			{Liu}},\ and\ \bibinfo {author} {\bibfnamefont {X.~F.}\ \bibnamefont {Han}},\
	}\bibfield  {title} {\bibinfo {title} {Spin-orbit torque switching in
			perpendicular {${\mathrm{Y}}_{3}{\mathrm{Fe}}_{5}{\mathrm{O}}_{12}$/Pt}
			bilayer},\ }\href {https://doi.org/10.1063/1.5098033} {\bibfield  {journal}
		{\bibinfo  {journal} {Applied Physics Letters}\ }\textbf {\bibinfo {volume}
			{114}},\ \bibinfo {pages} {192409} (\bibinfo {year} {2019})}\BibitemShut
	{NoStop}%
	\bibitem [{\citenamefont {Popova}\ \emph {et~al.}(2001)\citenamefont {Popova},
		\citenamefont {Keller}, \citenamefont {Gendron}, \citenamefont {Thomas},
		\citenamefont {Brianso}, \citenamefont {Guyot}, \citenamefont {Tessier},\
		and\ \citenamefont {Parkin}}]{Popova2001}%
	\BibitemOpen
	\bibfield  {author} {\bibinfo {author} {\bibfnamefont {E.}~\bibnamefont
			{Popova}}, \bibinfo {author} {\bibfnamefont {N.}~\bibnamefont {Keller}},
		\bibinfo {author} {\bibfnamefont {F.}~\bibnamefont {Gendron}}, \bibinfo
		{author} {\bibfnamefont {L.}~\bibnamefont {Thomas}}, \bibinfo {author}
		{\bibfnamefont {M.-C.}\ \bibnamefont {Brianso}}, \bibinfo {author}
		{\bibfnamefont {M.}~\bibnamefont {Guyot}}, \bibinfo {author} {\bibfnamefont
			{M.}~\bibnamefont {Tessier}},\ and\ \bibinfo {author} {\bibfnamefont
			{S.~S.~P.}\ \bibnamefont {Parkin}},\ }\bibfield  {title} {\bibinfo {title}
		{Perpendicular magnetic anisotropy in ultrathin yttrium iron garnet films
			prepared by pulsed laser deposition technique},\ }\href
	{https://doi.org/10.1116/1.1392395} {\bibfield  {journal} {\bibinfo
			{journal} {Journal of Vacuum Science \& Technology A}\ }\textbf {\bibinfo
			{volume} {19}},\ \bibinfo {pages} {2567} (\bibinfo {year}
		{2001})}\BibitemShut {NoStop}%
	\bibitem [{\citenamefont {Maier-Flaig}\ \emph {et~al.}(2018)\citenamefont
		{Maier-Flaig}, \citenamefont {Goennenwein}, \citenamefont {Ohshima},
		\citenamefont {Shiraishi}, \citenamefont {Gross}, \citenamefont {Huebl},\
		and\ \citenamefont {Weiler}}]{DerivativeDivide2018}%
	\BibitemOpen
	\bibfield  {author} {\bibinfo {author} {\bibfnamefont {H.}~\bibnamefont
			{Maier-Flaig}}, \bibinfo {author} {\bibfnamefont {S.~T.~B.}\ \bibnamefont
			{Goennenwein}}, \bibinfo {author} {\bibfnamefont {R.}~\bibnamefont
			{Ohshima}}, \bibinfo {author} {\bibfnamefont {M.}~\bibnamefont {Shiraishi}},
		\bibinfo {author} {\bibfnamefont {R.}~\bibnamefont {Gross}}, \bibinfo
		{author} {\bibfnamefont {H.}~\bibnamefont {Huebl}},\ and\ \bibinfo {author}
		{\bibfnamefont {M.}~\bibnamefont {Weiler}},\ }\bibfield  {title} {\bibinfo
		{title} {Note: Derivative divide, a method for the analysis of broadband
			ferromagnetic resonance in the frequency domain},\ }\href
	{https://doi.org/10.1063/1.5045135} {\bibfield  {journal} {\bibinfo
			{journal} {Review of Scientific Instruments}\ }\textbf {\bibinfo {volume}
			{89}},\ \bibinfo {pages} {076101} (\bibinfo {year} {2018})}\BibitemShut
	{NoStop}%
	\bibitem [{\citenamefont {Hillebrands}\ and\ \citenamefont
		{Thiaville}(2006)}]{HillbrandsBook}%
	\BibitemOpen
	\bibfield  {author} {\bibinfo {author} {\bibfnamefont {B.}~\bibnamefont
			{Hillebrands}}\ and\ \bibinfo {author} {\bibfnamefont {A.}~\bibnamefont
			{Thiaville}},\ }\href {https://doi.org/10.1007/b12462} {\emph {\bibinfo
			{title} {Spin Dynamics in Confined Magnetic Structures III}}}\ (\bibinfo
	{publisher} {Springer},\ \bibinfo {address} {Berlin, Heidelberg},\ \bibinfo
	{year} {2006})\BibitemShut {NoStop}%
	\bibitem [{\citenamefont {Collet}\ \emph {et~al.}(2016)\citenamefont {Collet},
		\citenamefont {De~Milly}, \citenamefont {Kelly}, \citenamefont {Naletov},
		\citenamefont {Bernard}, \citenamefont {Bortolotti}, \citenamefont {Youssef},
		\citenamefont {Demidov}, \citenamefont {Demokritov}, \citenamefont {Prieto},
		\citenamefont {Mu$\mathrm{\tilde{n}}$oz}, \citenamefont {Cros}, \citenamefont
		{Anane}, \citenamefont {de~Loubens},\ and\ \citenamefont
		{Klein}}]{collet2016generation}%
	\BibitemOpen
	\bibfield  {author} {\bibinfo {author} {\bibfnamefont {M.}~\bibnamefont
			{Collet}}, \bibinfo {author} {\bibfnamefont {X.}~\bibnamefont {De~Milly}},
		\bibinfo {author} {\bibfnamefont {O.~d.}\ \bibnamefont {Kelly}}, \bibinfo
		{author} {\bibfnamefont {V.~V.}\ \bibnamefont {Naletov}}, \bibinfo {author}
		{\bibfnamefont {R.}~\bibnamefont {Bernard}}, \bibinfo {author} {\bibfnamefont
			{P.}~\bibnamefont {Bortolotti}}, \bibinfo {author} {\bibfnamefont {J.~B.}\
			\bibnamefont {Youssef}}, \bibinfo {author} {\bibfnamefont {V.}~\bibnamefont
			{Demidov}}, \bibinfo {author} {\bibfnamefont {S.}~\bibnamefont {Demokritov}},
		\bibinfo {author} {\bibfnamefont {J.~L.}\ \bibnamefont {Prieto}}, \bibinfo
		{author} {\bibfnamefont {M.}~\bibnamefont {Mu$\mathrm{\tilde{n}}$oz}},
		\bibinfo {author} {\bibfnamefont {V.}~\bibnamefont {Cros}}, \bibinfo {author}
		{\bibfnamefont {A.}~\bibnamefont {Anane}}, \bibinfo {author} {\bibfnamefont
			{G.}~\bibnamefont {de~Loubens}},\ and\ \bibinfo {author} {\bibfnamefont
			{O.}~\bibnamefont {Klein}},\ }\bibfield  {title} {\bibinfo {title}
		{Generation of coherent spin-wave modes in yttrium iron garnet microdiscs by
			spin--orbit torque},\ }\href {https://doi.org/10.1038/ncomms10377} {\bibfield
		{journal} {\bibinfo  {journal} {Nature communications}\ }\textbf {\bibinfo
			{volume} {7}},\ \bibinfo {pages} {10377} (\bibinfo {year}
		{2016})}\BibitemShut {NoStop}%
	\bibitem [{\citenamefont {Cornelissen}\ \emph {et~al.}(2016)\citenamefont
		{Cornelissen}, \citenamefont {Peters}, \citenamefont {Bauer}, \citenamefont
		{Duine},\ and\ \citenamefont {van Wees}}]{LudoPRB}%
	\BibitemOpen
	\bibfield  {author} {\bibinfo {author} {\bibfnamefont {L.~J.}\ \bibnamefont
			{Cornelissen}}, \bibinfo {author} {\bibfnamefont {K.~J.~H.}\ \bibnamefont
			{Peters}}, \bibinfo {author} {\bibfnamefont {G.~E.~W.}\ \bibnamefont
			{Bauer}}, \bibinfo {author} {\bibfnamefont {R.~A.}\ \bibnamefont {Duine}},\
		and\ \bibinfo {author} {\bibfnamefont {B.~J.}\ \bibnamefont {van Wees}},\
	}\bibfield  {title} {\bibinfo {title} {Magnon spin transport driven by the
			magnon chemical potential in a magnetic insulator},\ }\href
	{https://doi.org/10.1103/PhysRevB.94.014412} {\bibfield  {journal} {\bibinfo
			{journal} {Physical Review B}\ }\textbf {\bibinfo {volume} {94}},\ \bibinfo
		{pages} {014412} (\bibinfo {year} {2016})}\BibitemShut {NoStop}%
	\bibitem [{\citenamefont {Takei}(2019)}]{Takei_Theory}%
	\BibitemOpen
	\bibfield  {author} {\bibinfo {author} {\bibfnamefont {S.}~\bibnamefont
			{Takei}},\ }\bibfield  {title} {\bibinfo {title} {Spin transport in an
			electrically driven magnon gas near bose-einstein condensation:
			Hartree-fock-keldysh theory},\ }\href
	{https://doi.org/10.1103/PhysRevB.100.134440} {\bibfield  {journal} {\bibinfo
			{journal} {Physical Review B}\ }\textbf {\bibinfo {volume} {100}},\ \bibinfo
		{pages} {134440} (\bibinfo {year} {2019})}\BibitemShut {NoStop}%
	\bibitem [{Note2()}]{Note2}%
	\BibitemOpen
	\bibinfo {note} {$\alpha _\protect \mathrm {eff}=\alpha _\protect \mathrm
		{G}+\delta H\left ( 2\protect \sqrt {H(H+M_\protect \mathrm {eff})}\right
		)^{-1}$}\BibitemShut {NoStop}%
	\bibitem [{Note3()}]{Note3}%
	\BibitemOpen
	\bibinfo {note} {$g_\protect \mathrm {eff}=\left [ {g^{\uparrow \downarrow }
			\protect \frac {h \sigma _\protect \mathrm {Pt}}{2e^2l_\protect \mathrm
				{s}}}\right ] / \left [ {g^{\uparrow \downarrow } + \protect \frac {h\sigma
				_\protect \mathrm {Pt}}{2e^2l_\protect \mathrm {s}}} \right ] $ and $\eta
		=t_\protect \mathrm {Pt}/(2l_\protect \mathrm {s})$}\BibitemShut {NoStop}%
	\bibitem [{\citenamefont {Althammer}\ \emph {et~al.}(2013)\citenamefont
		{Althammer}, \citenamefont {Meyer}, \citenamefont {Nakayama}, \citenamefont
		{Schreier}, \citenamefont {Altmannshofer}, \citenamefont {Weiler},
		\citenamefont {Huebl}, \citenamefont {Gepr\"ags}, \citenamefont {Opel},
		\citenamefont {Gross}, \citenamefont {Meier}, \citenamefont {Klewe},
		\citenamefont {Kuschel}, \citenamefont {Schmalhorst}, \citenamefont {Reiss},
		\citenamefont {Shen}, \citenamefont {Gupta}, \citenamefont {Chen},
		\citenamefont {Bauer}, \citenamefont {Saitoh},\ and\ \citenamefont
		{Goennenwein}}]{AlthammerSMR2013}%
	\BibitemOpen
	\bibfield  {author} {\bibinfo {author} {\bibfnamefont {M.}~\bibnamefont
			{Althammer}}, \bibinfo {author} {\bibfnamefont {S.}~\bibnamefont {Meyer}},
		\bibinfo {author} {\bibfnamefont {H.}~\bibnamefont {Nakayama}}, \bibinfo
		{author} {\bibfnamefont {M.}~\bibnamefont {Schreier}}, \bibinfo {author}
		{\bibfnamefont {S.}~\bibnamefont {Altmannshofer}}, \bibinfo {author}
		{\bibfnamefont {M.}~\bibnamefont {Weiler}}, \bibinfo {author} {\bibfnamefont
			{H.}~\bibnamefont {Huebl}}, \bibinfo {author} {\bibfnamefont
			{S.}~\bibnamefont {Gepr\"ags}}, \bibinfo {author} {\bibfnamefont
			{M.}~\bibnamefont {Opel}}, \bibinfo {author} {\bibfnamefont {R.}~\bibnamefont
			{Gross}}, \bibinfo {author} {\bibfnamefont {D.}~\bibnamefont {Meier}},
		\bibinfo {author} {\bibfnamefont {C.}~\bibnamefont {Klewe}}, \bibinfo
		{author} {\bibfnamefont {T.}~\bibnamefont {Kuschel}}, \bibinfo {author}
		{\bibfnamefont {J.-M.}\ \bibnamefont {Schmalhorst}}, \bibinfo {author}
		{\bibfnamefont {G.}~\bibnamefont {Reiss}}, \bibinfo {author} {\bibfnamefont
			{L.}~\bibnamefont {Shen}}, \bibinfo {author} {\bibfnamefont {A.}~\bibnamefont
			{Gupta}}, \bibinfo {author} {\bibfnamefont {Y.-T.}\ \bibnamefont {Chen}},
		\bibinfo {author} {\bibfnamefont {G.~E.~W.}\ \bibnamefont {Bauer}}, \bibinfo
		{author} {\bibfnamefont {E.}~\bibnamefont {Saitoh}},\ and\ \bibinfo {author}
		{\bibfnamefont {S.~T.~B.}\ \bibnamefont {Goennenwein}},\ }\bibfield  {title}
	{\bibinfo {title} {Quantitative study of the spin {Hall} magnetoresistance in
			ferromagnetic insulator/normal metal hybrids},\ }\href
	{https://doi.org/10.1103/PhysRevB.87.224401} {\bibfield  {journal} {\bibinfo
			{journal} {Physical Review B}\ }\textbf {\bibinfo {volume} {87}},\ \bibinfo
		{pages} {224401} (\bibinfo {year} {2013})}\BibitemShut {NoStop}%
	\bibitem [{\citenamefont {Zhang}\ \emph {et~al.}(2015)\citenamefont {Zhang},
		\citenamefont {Han}, \citenamefont {Jiang}, \citenamefont {Yang},\ and\
		\citenamefont {Parkin}}]{ZhangInterfaceTransparency}%
	\BibitemOpen
	\bibfield  {author} {\bibinfo {author} {\bibfnamefont {W.}~\bibnamefont
			{Zhang}}, \bibinfo {author} {\bibfnamefont {W.}~\bibnamefont {Han}}, \bibinfo
		{author} {\bibfnamefont {X.}~\bibnamefont {Jiang}}, \bibinfo {author}
		{\bibfnamefont {S.-H.}\ \bibnamefont {Yang}},\ and\ \bibinfo {author}
		{\bibfnamefont {S.}~\bibnamefont {Parkin}},\ }\bibfield  {title} {\bibinfo
		{title} {Role of transparency of platinum-ferromagnet interface in
			determining intrinsic magnitude of spin hall effect},\ }\href
	{https://doi.org/10.1038/nphys3304} {\bibfield  {journal} {\bibinfo
			{journal} {Nature Physics}\ }\textbf {\bibinfo {volume} {11}},\ \bibinfo
		{pages} {496} (\bibinfo {year} {2015})}\BibitemShut {NoStop}%
\end{thebibliography}
\end{document}